# Comment to the Review by L. Glozman, hep-ph/0701081

Recently Glozman has posted a review to the ArXiv, which is announced to be published in Physics Reports [1]. This review is based on about seven years of the author's original research on chiral properties of the hadron spectra, as well as contributions of other experts. The author [1] reviewed the developments on a very intriguing topic, i.e. restoration of the chiral symmetry and $U(1)_A$ symmetry in excited hadrons, considering it in a framework of the well-known models and methods on a market – string model, operator product expansion method (OPE), generalized linear sigma model, 't Hooft model, WKB-approximation, large $N_c$ method, and others.

Nevertheless, I have to admit that many of the author's [1] comments, claims, ideas, and proofs are far from being accepted under real scrutiny and have to be revised. Let us consider one of the central issues in the review [1], namely "two-regimes" or "two-phases" scheme for the chiral hadron dynamics. Throughout all of the paper, Glozman claimed – "at low quark's momentum chiral symmetry is strongly broken by the vacuum, and at high quark's momentum the chiral symmetry is restored, as vacuum effects vanished". An evident question arises here – what is low quark's momentum (or energy), and high quark's momentum (or energy) were meant by the author? He had never specified it, or gave estimates based on some strong interaction dynamics!

I think most of the results [1] crucially depend in one or the other ways on these parameter's values. What's more, why the author [1] even has decided that we are dealing with the "two-phase" chiral structure of excited hadrons? He gave neither explanation, nor proof for that matter. Recently Afonin clearly shows [2] that "The physics above $\Lambda_{CSR}$ is indistinguishable from the physics of perturbative QCD continuum. The scale $\Lambda_{CSR}$ marks transition from intermediate to high energy like the scale $\Lambda_{CSB}$ does from low to the intermediate one. Thus, if experimentally confirmed, the scale $\Lambda_{CSR}$ is of great importance because it marks the upper bound of resonance physics in the light quark sector of QCD" – so, we clearly see that Afonin implied the natural existence of the broad intermediate region with distinctive physics.

The author [1] wrote on the same topic – "Members of the given parity-chiral multiplet, become close at large s (and identical asymptotically high)". What does it means precisely – asymptotically high? Asymptotically, baryon and meson states will create a continuum, so there should be a reasonable estimate for a "threshold s-value", corresponding to the transition from pre- to the full asymptotics! We also have 3 chiral regimes in our recent paper [3]. Kalashnikova et al [4] in GNJL model also clearly had showed 3 regions in a mesonic structure:

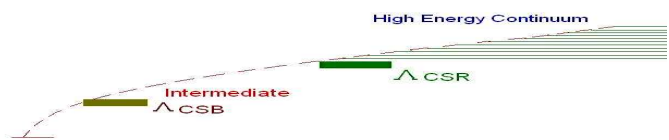

One of the central topics in review [1] is a string model of hadrons and its role in forming the big picture of the chiral symmetry restoration (Section 12). Glozman's line of reasoning had started with "paradigm" – "quarks are massless, hence they move at the speed of light and (hence) have some *fixed chirality* ". So far so good. But then the author [1] stepped forward with massive quarks, and transferred all the conclusions about *fixed chirality* to this situation. The real things are: string with massless ends is only the simplest model, and it doesn't describe spins at all. To describe quark spins we are to consider quarks with masses (string with massive ends). These masses move at speeds v < c. So, the further conclusions about fixed chirality are wrong. Right after that, author [1] comes quickly to the "conclusions" about vanishing of the spin-orbit and tensor forces in a string hadron scenario. These conclusions are based on the wrong statement about chirality – for massive quarks the velocity vector is not uniquely preferred, so the spin projections onto z-axis (perpendicular to the rotational plane) may be nonzero:

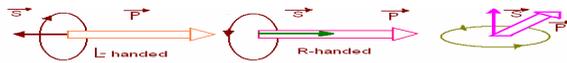

Actually, there are much more configurations possible for 3q-strings, than author [1] has considered. Many of them have very distinctive topologies, not considered in this review at al. The author completely ignores different roles for different configurations in the string-hadron picture. Each of them contributed differently to the energy- and spin- properties of the given hadron [5]. Especially interesting are "exotic" configurations [5], which might be applicable for the glueball structures.

Recently 't Hooft has made an interesting observation: "As soon as classical string picture is adopted for baryonic states, at least one of three arms will soon *disappea*r, shedding its energy into excitation modes of the two other arms [6]", see also there [6]: "For RT's, at any given energy, highest angular momentum states will be achieved when one arm vanishes" – nothing like that can be seen in review [1]

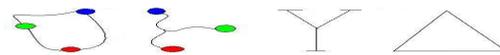

Again, 't Hooft reads: " If, due to gluon hitting a quark, a baryonic state is created with 3 quarks energetically moving in different directions, we expect first the Y shape to form , but then most likely baryonic excitation that is reached is one with a single open string connecting 3 quarks" [6]. Sadly, but we have to conclude that author [1] has confused everything possible in a string dynamics of hadrons.

One of the major drawbacks in review [1] is an almost total absence of the discussion of the $\Lambda$ - $\Sigma$ baryonic sector. There are plenty of interesting chiral physics in $\Lambda$ - $\Sigma$ and this was clearly shown by us [7], and also in Klempt's [8], and Kirchbach's papers [9]. What's even more, Glozman didn't realize the full symmetrical circle of the $N$ - $\Delta$ - $\Lambda$ - $\Sigma$ and it's potential to be described by one isotopical superstate with new seashell multiplets: $I = 0 + 1/2 + 1 + 3/2$. In parity doublets we have a generalization from in–plane $N$ - $\Delta$ rotator [10] to the 3D cylinder with $\Lambda$ - $\Sigma$ on top floor and $N$ - $\Delta$ on the first floor. Jaffe also has investigated recently $N$ - $\Delta$ and $\Lambda$ - $\Sigma$ sectors together and found some clear evidence for the creation of parity doublers in both sectors [11].

It was exactly due to the Klempt's model [8] that we have generated new $N$ - $\Delta$ - $\Lambda$ - $\Sigma$ supermultiplets [7] :

$$L=1: \quad N \to \Sigma \to \Lambda \to \Delta$$

$$L=2: \quad \Delta \to \Sigma \to \Lambda \to N$$

$$L=3: \quad N \to \Delta$$

$$L=4: \quad \Delta$$

Now, just look at the structure of our supermultiplets [7], emulated from the Table5: $L=1$ supermultiplet is exactly {$\Lambda\Sigma$, $N\Delta$} super seashell, which embraces all four species together: $\Lambda$ - $\Sigma$ - $N$ - $\Delta$. The $L=2$ floor gives the picture of the two "ordinary" seashells {$\Delta$ - $N$} and {$\Lambda$ - $\Sigma$}.

Let's return to our paper [7] and continue with all parallels between {$N$-$\Delta$}, {$\Lambda$-$\Sigma$}, and general {$N$-$\Delta$-$\Lambda$-$\Sigma$} seashells to elicit the hidden dynamics and symmetries. In the Cohen-Glozman original chiral multiplets (CG) we have two seashells with $J=3/2$ and $J=7/2$. But if we consider physically more sound Cohen-Glozman-Inopin (CGI) chiral multiplets, we will find already four seashells with $J=1/2, 7/2, 15/2$, and $19/2$. We can create even more order among the {$N$-$\Delta$} seashells. One can see the following "$N$ in $\Delta$" seashells: $J=7/2, J=15/2$, and $J=19/2$. From the other hand we have the following "$\Delta$ in $N$" seashells: $J=1/2$ and $J=3/2$ only. There is a clear pattern – "$\Delta$ in $N$" seashells exist only for a small $J=1/2, 3/2$, and "$N$ in $\Delta$" seashells exist only for higher spins: $J=7/2, 15/2, 19/2$. I think this phenomenon is reflecting some deep symmetrization properties of the $N$-$\Delta$ chiral multiplets.

Now, what is a physical meaning for all the rest of $N$-$\Delta$ chiral multiplets? They are playing the role of the intermediate configurations, tending to become either perfect seashells, or the "solitaire" parity doublets. Let's be more precise: $J=5/2, 9/2$ consists of the two widely separated $N$ and $\Delta$ parity doublets, $J=11/2, 13/2, 17/2$ are $N$-$\Delta$ overlapped quartets (but $J=17/2$ could be easily attributed to seashell), and $J=1/2, 3/2, 7/2, 15/2, 17/2, 19/2$ are pure seashells. One can see that pure seashells are to be realized in the most of the configurations – 6 out of 10, i.e. 60%.

Finally, let's turn to the estimates which Glozman made [1] for the chiral asymmetry $\chi$, and overlap $S_0$, which characterize the restoration of chiral symmetry in different hadronic multiplets. Because of the large bias in mesonic multiplet's assignments, we will compute these parameters for the different baryonic chiral multiplets from our paper [7]. We started from an original Cohen-Glozman (CG) $N$ - $\Delta$ chiral multiplets, and obtained interesting results: for $J= 1/2, 3/2, 5/2, 7/2, 9/2$ the chiral asymmetry $\chi$ is much bigger than value 0.02, claimed in [1], with average value $\chi = 0.053$, or 5.3 %. Let's analyze situation with overlap parameter $S_0$ for CG multiplets. Overlaps typically happened to be much bigger than an estimate 0.2, suggested by the author [1]. Only $S_0(9/2) = 0.83 <1$, and all the rest are much bigger with $S_0$ average = 2.92 >> 0.2 !
So, we refute author's [1] claim about $\chi$, $S_0$ smallness for his own chiral multiplets.

Now let us do the same exercise for the Cohen-Glozman-Inopin (CGI) $N$-$\Delta$ chiral multiplets [7]. We have much more multiplets there, from $J=1/2$ all the way to $J =21/2$, and the parameter $\chi$ is wildly fluctuating from 0.0013 to 0.088, with mostly much bigger than suggested 0.02 value, and $\chi$ average = 0.052. For the CGI overlap $S_0$ we have clear sequence with all of $S_0 >1$, except $S_0(3/2) = 0.47$, and with $S_0$ average = 2.36, which is 11 times the naive estimate from [1]. We have proved the failure of Glozman's claims about $\chi$ and $S_0$ values for two sets of $N - \Delta$ chiral multiplets. In the Table 3, p.32 the author [1] has fixed the spacings between consecutive chiral multiplets at the same value 200 MeV. The direct comparison with Cohen – Glozman original multiplets [7] gives the set of values: 40 MeV, 49 MeV, 74 MeV, 210 MeV, and 217 MeV with average spacing being 118 MeV. Again, one can see how rough the numerical work has been done in [1].

In no way we have tried here to refute or reconsider all the claims, conclusions and proofs from the review [1] – there are too numerous to list in a short Comment. We would suggest to the author to seriously revisit this paper, before sending it to the Physical Reports.

A. E. Inopin
Department of Experimental Nuclear Physics
Kharkov National University
Svobody Sq. 4, 61077, Kharkov, Ukraine